\documentclass[prl,
                twocolumn,
                preprintnumbers,
                amsmath,
                amssymb,
                showpacs,
                superscriptaddress]{revtex4-1}
\usepackage{graphicx} 
\usepackage{dcolumn}  
\usepackage{bm}       
\usepackage{amsfonts}

\usepackage{amssymb}
\usepackage[dvipsnames]{xcolor}
\usepackage{soul}
\usepackage{braket}
\usepackage{amsmath}
\usepackage[normalem]{ulem}

\newcommand{\eq}[1]{Eq.~(\ref{#1})}
\newcommand{\na}{\mbox{Na$_{20}$}}
\newcommand{\ful}{\mbox{C$_{240}$}}
\newcommand{\naful}{\mbox{Na$_{20}$@C$_{240}$}}

\begin{document}

\title{Plasmonic Resonant Intercluster Coulombic Decay}

\author{Rasheed Shaik}
\affiliation{%
School of Physical Sciences, Indian Institute of Technology Mandi, Kamand, H.P. 175075, India}

\author{Hari R. Varma}
\email[]{hari@iitmandi.ac.in}
\affiliation{%
School of Physical Sciences, Indian Institute of Technology Mandi, Kamand, H.P. 175075, India}

\author{Mohamed El-Amine Madjet}
\affiliation{Bremen Center for Computational Materials Science, University of Bremen, Bremen, Germany}
\affiliation{Department of Natural Sciences, D.L.\ Hubbard Center for Innovation, Loess Hills Research Center, Northwest Missouri State University, Maryville, Missouri 64468, USA}

\author{Fulu Zheng}
\affiliation{Bremen Center for Computational Materials Science, University of Bremen, Bremen, Germany}

\author{Thomas Frauenheim}
\affiliation{Bremen Center for Computational Materials Science, University of Bremen, Bremen, Germany}
\affiliation{Beijing Computational Science Research Center, 100193 Beijing, China}
\affiliation{Shenzhen JL Computational Science and Applied Research Institute, 518110 Shenzhen, China}

\author{Himadri S. Chakraborty}
\email[]{himadri@nwmissouri.edu}
\affiliation{Department of Natural Sciences, D.L.\ Hubbard Center for Innovation, Loess Hills Research Center, Northwest Missouri State University, Maryville, Missouri 64468, USA}

\begin{abstract}
{Light-induced energy confinement in nanoclusters via plasmon excitations influences applications in nanophotonics, photocatalysis, and the design of controlled slow electron sources. The resonant decay of these excitations through the cluster's ionization continuum provides a unique probe of the collective electronic behavior. However, the transfer of a part of this decay amplitude to the continuum of a second conjugated cluster may offer control and efficacy in sharing the energy nonlocally to instigate remote collective events. With the example of a spherically nested dimer $\naful$ of two plasmonic systems we find that such a transfer is possible through the resonant intercluster Coulomb decay (RICD) as a fundamental process. This plasmonic RICD signal can be experimentally detected by the photoelectron velocity map imaging technique.}

\end{abstract}

\maketitle
Resonant (electronic) energy transfer (RET), first observed a century ago~\cite{cairo1922}, is mediated by virtual photon exchange between weakly bonded sites embedded in chemical or biological environments. Both donor and acceptor sites are generally called chromophores, with clean absorption and fluorescence bands, coupled predominantly by the quantum electrodynamic dipole interaction~\cite{jones2019}. Fast forward to the present and the RET studies involving nanomaterials have flourished with attractive applications. For instance, there are bio-inspired RET processes using quantum dots to perform photodynamic cancer therapy~\cite{li2012}, bio- and nanosensor energy harvesting~\cite{stanisavljevic2015,hildebrandt2017}, and many others~\cite{liu2015,jang2004}. RET pathways involving carbon nanotubes are also researched~\cite{mhlenbacher2015,davoody2016}. Particularly fascinating are the processes {\em mediated} by the plasmon excitation, the collective excitation of conduction electrons. Owing to the efficient, large-scale energy containment and flow abilities of these excitations, the plasmon-liaised RET raised many interests~\cite{hsu2017,torres2016}. However, the RET processes are confined in the excitations of the upper-lying electron levels that occur within the visible to mid-ultraviolet electromagnetic spectrum. 

A different class of processes emerges when the absorbed photon is within the range of extreme ultraviolet (XUV) to x-ray. They induce innershell excitations in donors but outershell ionization in acceptors with the Coulomb interaction between the excited and the ionized electrons being the intermediary of the process. These interatomic and intermolecular Coulombic decay (ICD) processes have been the subject of a vast range of studies~\cite{jahnke2020} since its discovery by Cederbaum and collaborators~\cite{cederbaum1997} more than two decades ago. In one original experiment, the precursor excitation was induced in Ne dimers by the synchrotron radiation~\cite{jahnke2004rareDimer}. Later for higher pulse rates to carry out time-resolved measurements, free electron laser sources are found appropriate~\cite{schnorr2013}. On the other hand, ICD signatures are also probed by electron~\cite{marburger2003firstExp} and ion~\cite{wiegandt2019} spectroscopy. These include various coincidence techniques, namely, the velocity map imaging (VMI) technique~\cite{laforge2019}. Contemporary pump-probe approaches~\cite{schnorr2013,takanashi2017}, specifically by light field streaking techniques~\cite{trinter2013}, to access time-resolved ICD are also made possible.  Recently, the measurement of ICD in liquid water to draw comparisons with ICD in water clusters~\cite{zhang2021} and the prediction about the control of ICD in the cavity of quantum light~\cite{cederbaum2021} have been published. ICD electrons for unbound (gaseous) system of pyridine monomers based on the energy-transfer through associative interactions has just been measured~\cite{barik2022}.  
\begin{figure*}[t]
\includegraphics[width=0.9\textwidth]{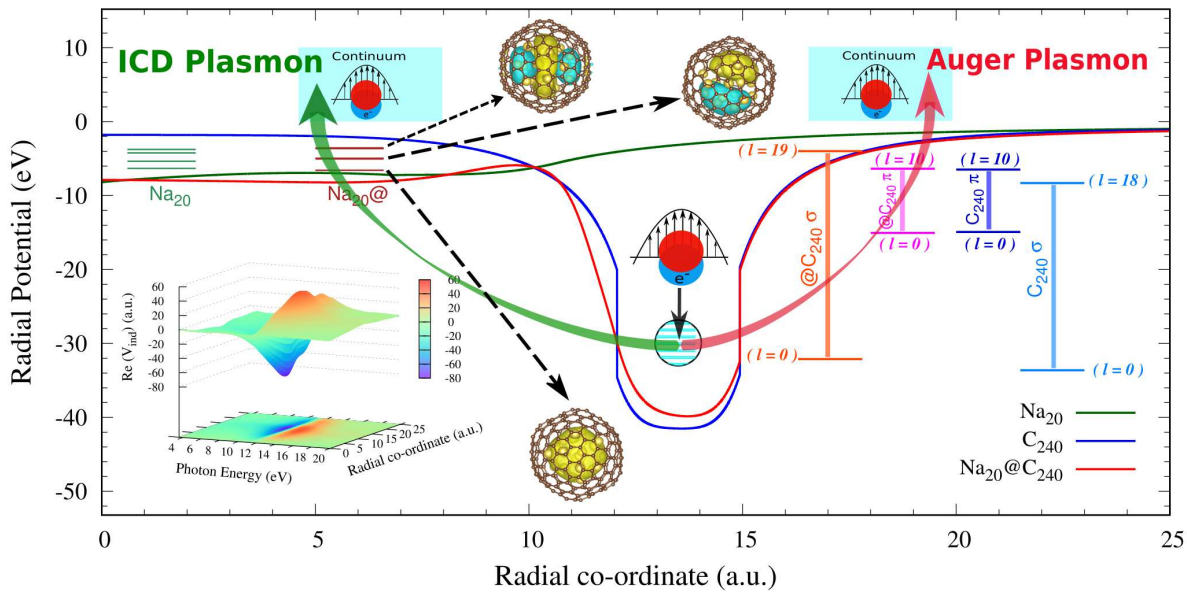}
\caption{(Color online) A cartoon to delineate the process of resonant Auger and ICD plasmon decays in $\naful$. The DFT radial potentials of $\naful$, isolated $\na$, and empty $\ful$ are drawn (solid curves). The corresponding $\na$ occupied energy levels as well as $\ful$ $\pi$ and $\sigma$ occupied band edges, defined by the maximum and minimum angular quantum number ($l$), are shown; among the levels of $\naful$, the symbols $\na$@ and @$\ful$ are used to distinguish from free system levels. The isosurface orbitals (HOMO, HOMO-6, HOMO-23) from quantum chemical calculations representing the three $\na$@ states are shown. Inset: The real part of the LR-TDDFT induced radial potential for $\naful$.}
\label{fig1}
\end{figure*}

Remarkably, no evidence of the decay of a plasmon resonance through an ICD channel is yet found. However, owing to characteristic extreme light confinement and control abilities within the nanoscale size, the plasmonic ICD will not only be a fundamental process, but also can substantially enrich the scopes within the vast ICD landscape and related applications in physics, chemistry, and biology. For instance, possibility and control of plasmon-driven remote-photonics, remote-catalysis or even remote-release of secondary electrons can be within the technological reach. With this motivation, we found an efficient prototype system in the nanocluster $\naful$ to probe the {\em intercluster} Coulombic decay of a plasmon excitation and revealed the first evidence of this fundamental phenomenon. 

Structure studies~\cite{trujillo1996} of $\naful$ and a recent review~\cite{yang2017} on general clusterfullerenes are available. $\naful$ is a spherical compound of a sodium sub-nano particle endohedrally confined in a carbon fullerene forming a nanometric spherical dimer. Both the units support their intrinsic (native) plasmon excitations. The plasmon of $\na$ excites in the visible spectrum below its first ionization threshold in the absorption response~\cite{xia2009,madjet2009}, while the giant plasmon (GP) excitation of $\ful$ embeds in the ionization continuum at the XUV spectral region, similar to the well known GP of C$_{60}$. Being energetically so separated, the hybridization~\cite{mccune2011} of these native plasmons is forbidden in $\naful$ ensuring reliability of the current result. Incidentally, C$_{60}$ with confined atoms are lately studied for variants of ICD processes~\cite{javani2014,magrakvelidze2016,de2016,khokhlova2020,de2021}, but for ordinary single-electron vacancy decay resonances. An excellent recent experiment to measure the ICD relaxation between the holmium nitride molecule and its C$_{80}$ cage has been reported~\cite{obaid2020}. 

The details of the computational methodology employed are given as supplementary materials (SM)~\cite{SM-method}. The ground states (1$s^2$1$p^6$1$d^{10}$2$s^2$ in the harmonic oscillator notation) of the free $\na$ cluster, the empty $\ful$ molecule, and the endofullerene $\naful$ are modeled by a jellium-based density functional theory (DFT) in the spherical frame. The DFT ground configuration of $\naful$ well replicates our detailed quantum chemical (QC) calculations based on the Turbomole software with the B3LYP exchange-correlation functional providing significant accuracy of the ground state description. This entailed the transfer of six $\na$ electrons to the fullerene shell leaving fourteen valence delocalized electrons (1$s^2$1$p^6$1$d^{6}$) in the cluster. The DFT radial potentials are plotted in Figure 1 along with the corresponding electronic energy-levels. As seen, the three $\na$ levels denoted by $\na$@ slightly shift in the compound; for details, see Figure S1 in SM~\cite{SM-method}. However, remarkably, the $d$, $p$, and $s$-type angular momentum character of these delocalized DFT $\na$@ levels are almost exactly reproduced by our simulated QC orbitals as shown. The fullerene bands of the $\pi$ (the group of one radial node) and $\sigma$ (the group of no radial node) states are identified in Fig.\,1 with band edges defined by the angular quantum number ($l$). The $\pi$ band exhibits stronger energy-stability with (@$\ful$) or without ($\ful$) $\na$. 

The dipole response of the systems to the incoming photon is described and computed by a linear response time-dependent DFT (LR-TDDFT) approach~\cite{choi2017,shaik-tobe}, since it is extremely challenging to use the QC framework to describe the electron continuum. On the other hand, the current approach has a track record of success in explaining measurements of (i) plasmonic photodepletion spectra~\cite{xia2009,madjet2009}, (ii) photoelectron~\cite{ruedel2002} and photoion~\cite{scully2005} intensity, respectively, at non-plasmonic and plasmonic energies, and (iii) plasmonic time delay spectra~\cite{biswas2022}. LR-TDDFT encodes the electron many-body correlations in the calculated induced potential, which is a complex quantity~\cite{SM-method}. The large-scale coherent component of the correlation over the GP formation energies blocks (screens) the incoming radiation but allows (anti-screens) the radiation to couple to the electrons at energies above the GP peak $\sim$ 11.5 eV. Indeed, this shape reversal is seen in Fig.\,1 (inset) in the real part of the induced radial potential for $\naful$ over the $\ful$ shell region that sluices through a zero at the GP peak energy.
\begin{figure}[h!]
\includegraphics[width=8.5 cm]{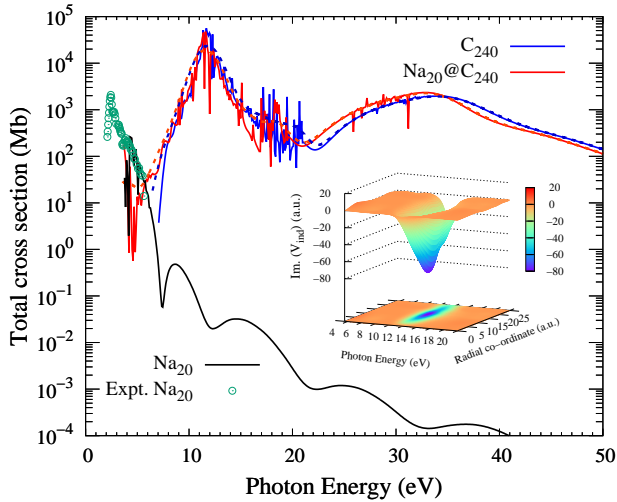}
\caption{LR-TDDFT total photoionization cross sections for $\ful$, $\naful$ and $\na$. Smoothed curves are added to guide the eye for the fullerene systems. The experimental absorption data~\cite{xia2009} are included for $\na$. Inset: The imaginary part of the induced radial potential for $\naful$ calculated in LR-TDDFT.}
\label{fig2}
\end{figure}

An elegant way to visualize the plasmon formation is to consider, in terms of many-body ground $|\Phi_0\rangle$ and excited collective $|\Phi_m\rangle$ states, the complex polarizability $\alpha$ from electrons' interactions with the photon of frequency $\omega$~\cite{zangwill80}. Using the Fermi's Golden rule, the absorption cross section $\sigma$ relates to the imaginary part of $\alpha$ as~\cite{madjet2008} 
\begin{eqnarray}\label{cross-pa}
\sigma (\omega) &\sim& \mbox{Im}[\alpha(\omega)] \nonumber \\
       \sim && \sum_m \left[\frac{|\langle\Phi_m|\zeta|\Phi_0\rangle|^2}
    {[\hbar\omega-\Delta_m]^2+\delta^2}-\frac{|\langle\Phi_m|\zeta|\Phi_0\rangle|^2}
       {[\hbar\omega+\Delta_m]^2+\delta^2}\right]
\end{eqnarray}
where $\zeta = \sum_i z_i$ are dipole interactions, $\Delta_m = E_m - E_0$ are many-body excitation energies, and $\delta$ is an infinitesimal positive quantity. \eq{cross-pa} embodies the notion that the plasmons at photon energies $\Delta_m$ are due
to excitations of the ground state $\Phi_0$ to all possible collective excited states $\Phi_m$ that the system supports.

Both the GP and a much weaker high-energy plasmon (HP) excitations of $\ful$ energetically embed in the system's ionization continuum. Thus, they produce GP resonance (GPR) and HPR in LR-TDDFT photoionization cross sections, as seen in Figure 2, both for $\ful$ and $\naful$. The narrow structures represent the incoherent single electron inner-shell excitation resonances and are not relevant for the current study. An attractive-shaped imaginary part of the induced radial potential of $\naful$ (inset of Fig.\,2), dipping at the GP peak and locating across the $\ful$ shell, transiently binds the collective excitation. An excellent agreement of the current theory with recent pump-probe streaking measurements accesses the resulting photoemission time delay at GP in C$_{60}$~\cite{biswas2022}. Both the plasmon resonances in $\ful$ can therefore be visualized as the decays through $\ful$ continuum emissions following the de-excitations of the plasmon states. In this spirit, they can be called the {\em Auger} plasmon resonances (Fig.\,1). However, the free $\na$ photoionization spectrum over this XUV energy range is very weak and completely plasmon-barren (Fig.\,2). In fact, $\na$'s native plasmon resonance excites at a far lower 2 eV photon energy as seen in the measured data~\cite{xia2009}; note also the agreement of the data at higher energies with the trend of the $\na$ cross section. Furthermore, the real (Fig.\,1) and the imaginary (Fig.\,2) parts of the induced radial potential are featureless over the central $\na$ region of the compound. Yet, our current results exhibit a plasmon-like resonance in the $\na$@ ionization channel which exhausts comparable oscillator strength (OS) used by the native plasmon resonance of $\na$ (see Figure 4). This must be the resonant ICD (RICD) transfer of the $\ful$ GP as schematically shown in Fig.\,1. Since the precursor plasmon excitation itself decays, the process qualifies as a participant RICD. The ICD transfer from HP is, however, negligibly weak and will be disregarded. 

A model-framework to understand this RICD transfer from the $\ful$ GP to $\na$ can be motivated based on an interchannel-coupling analysis following the well-known discrete-continuum coupling approach by Fano~\cite{fano1961}. The RICD amplitude of the $\ful$ plasmon ``vacancy" decay {\em via} the $\na$ $nl@$ ionization can be expressed by $M^{\mbox{\scriptsize p-c}}$. This will include the coupling of $\ful$ $0\rightarrow m$ plasmon (p) excitation channel with the $nl@\rightarrow kl'$ continuum (c) channel of $\na$. Hence, $M^{\mbox{\scriptsize p-c}}$ can be written as:
\begin{eqnarray}\label{pc-mat-element}
 {M}^{\mbox{\scriptsize p-c}}_{nl@}(E) &&\nonumber \\
 &&\sim \frac{\langle{0\rightarrow m}|\frac{1}{|{\bf r}_p-{\bf r}_{nl@}|}|{nl@\rightarrow kl'}(E)\rangle}{E-\Delta_{m}} {\cal D}_{0 \rightarrow m}
\end{eqnarray}
where ${\cal D}_{0 \rightarrow m}$ is the plasmonic amplitude of @$\ful$ photoionization and $E$ is the photon energy that enables the $\na$ $nl@$ transition to the continuum. By motivating ${\bf r}_p$ to be the ``co-ordinate" of the plasmon quasi-particle, the Coulomb-type coupling matrix element in the numerator of \eq{pc-mat-element} acts as the passage for virtual-energy transfer from the GP de-excitation across to the $\na$ ionization, producing plasmonic ICD resonances in a $\na$ $nl@$ cross section. In effect, a conduit of this passage is the overlap between the wave functions involving the collective excitation and the $\na$ ionization channels in the Coulomb matrix element.
\begin{figure*}[t]
\includegraphics[width=0.9\textwidth]{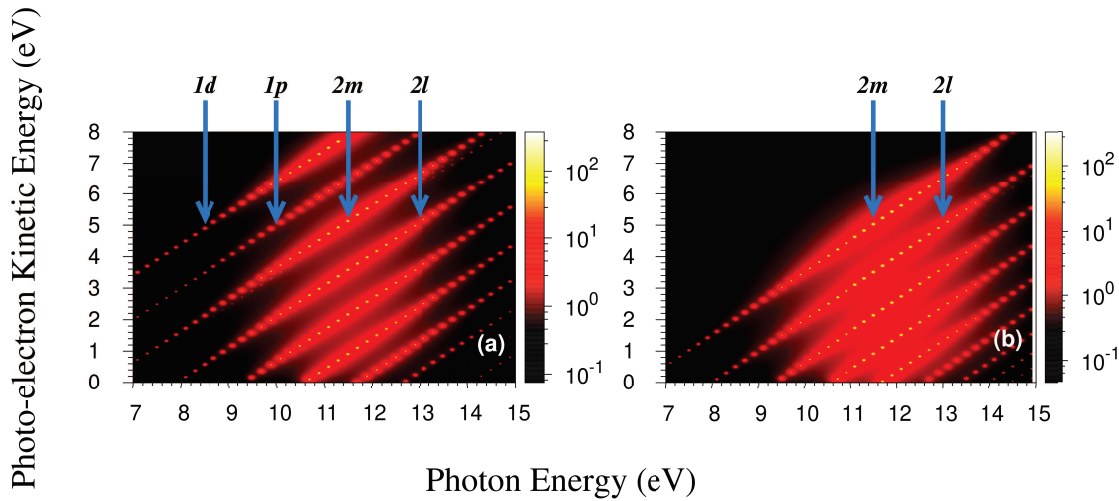}
\caption{(Color online) Color coded cross section maps as a function of photon energy and photoelectron kinetic energy for $\naful$ (a) and $\ful$ (b). The smoothed curves are used to solely capture the plasmon resonance signals and some blur is introduced to highlight the resonance profiles. Plasmonic RICD traces from 1$d$ and 1$p$ $\na$@ levels are identified in panel (a) (1$s$ is too weak to show in this scale), while both panels show similar distributions of regular Auger plasmon traces in $\ful$ levels within the range shown. Auger traces in both (a) and (b) are somewhat scaled down for better contrast.}
\label{fig3}
\end{figure*}

We capture the RICD signals in the $\na$@ channels by generating iso-contour images of cross section as a function of photon energy and photoelectron kinetic energy in Figure 3. Such images can be produced in the experiment by using the standard VMI coincidence technique~\cite{basnayake2022} to separate and selectively detect photoions or photoelectrons by mass and velocity. Here we used the cross sections after smoothing the profiles for narrow resonance spikes in order to feature only the broad plasmonic contributions. The vertical sections of the images provide the photoelectron energy distribution of the signal by mapping the ionization thresholds of the levels. The subshell cross sections as a function of the photon energy, on the other hand, can be extracted from horizontal sections. We now compare the image of Fig.\,3(a) for $\naful$ with Fig.\,3(b) for $\ful$. Obviously, the additional traces in Fig.\,3(a) are the $\na$ RICD signals, while the $\ful$ Auger signals are common in both image panels. Note, as expected, that all the traces peak within 11--12 eV photon energy range. 
\begin{figure}
\includegraphics[width=8.5 cm, keepaspectratio]{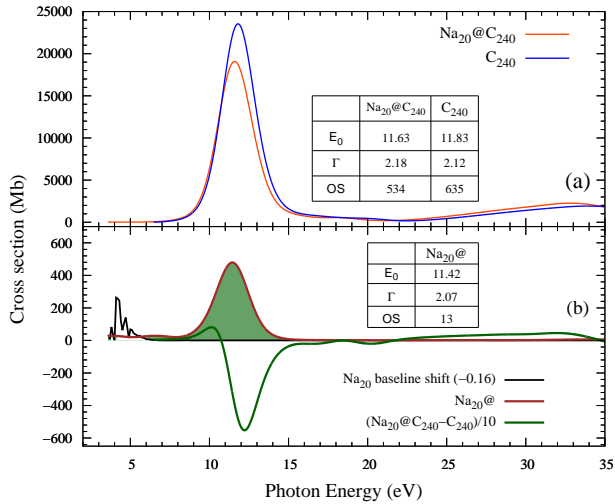}
\caption{(Color online) All smoothed LR-TDDFT cross section results are used in this figure. (a): The total cross sections for $\naful$ and $\ful$ are shown. The values of the fitting parameters of the peak energy (E$_0$) and the width ($\Gamma$, as well as the oscillator strength (OS) exhausted by the GPR are given in the table. (b): The difference (scaled-down) between $\naful$ and $\ful$ results, the raw $\na$ result (with a slight down-shift for the aid of comparison), and the RICD plasmon feature ($\na$@) are presented. The fitting parameters and the OS of the later are tabulated.}
\label{fig4}
\end{figure}

Fig.\,4(a) displays the total (smoothed) cross sections of $\naful$ and $\ful$. The curves are fitted with two Lorentzian profiles, accounting the two plasmon resonances, for characterization. The values of the fitting parameters for GPR are tabulated in the figure. The peak energy (E$_0$) is found to very slightly red-shift when $\na$ is present. The corresponding increase in the resonance line-width ($\Gamma$) from 2.12 eV to 2.18 eV amounts to somewhat shortening of the lifetime from 610 attoseconds (as) to 595 as. Going from $\ful$ to $\naful$, this effect must be due to the additional (ICD) decay rate introduced by the $\na$ ionization channels. The ICD contribution is shown independently as $\na$@ in Fig.\,4(b). This RICD plasmon feature yields a narrower line-width of 2.07 eV corresponding to a longer lifetime of about 625 as. The implication of this result is that the average dephasing of the RICD feature is slower than the resonant Auger feature -- a fact which may benefit time domain measurements to separate and probe the plasmonic ICD. The difference between $\naful$ and $\ful$ results in Fig.\,4(b) provides an energy-differential effect of the $\na$ doping at GPR energies. As seen, the doping results in a slight constructive interference at lower energies but a larger destructive interference at higher energies. In effect, there is a net reduction, which is reflected in the corresponding reduction of OS [Fig.\,4(a)] utilized by the resonances. However, the value of OS spent by the ICD plasmon resonance is found to be 13 [Fig.\,4(b)], which is remarkably close both to the OS ($\sim$15) consumed by the native plasmon resonance of $\na$~\cite{shaik2021} and the electron population of 14 at $\na$ in the compound. 

Making larger fullerenes are difficult. Even if achieved, the production may suffer from lower yields and higher isomer counts.  However, there is a recent approach of isolating larger fullerenes from fullertube isomers~\cite{koenig2020}. Also, among various techniques~\cite{popov2017} of synthesis and extraction of endofullerenes, the irradiation of fullerene film with metal ion beams~\cite{lee2020} is fairly successful. This method can be extended for longer time exposure to implant multiple ions to produce clustered encapsulation. However, the technique may deplete the sample yield from film destruction. We hope that the technology will improve so the future experiments can access the current pilot prediction of plasmonic RICD.

To conclude, using a many-body framework of DFT, very well-supported by a QC calculation of the ground state, we predict a hitherto unknown ICD dynamics of a plasmon resonance. It is shown that the XUV-photon driven giant plasmon of the $\ful$ fullerene can efficiently transfer strength via the resonant ICD through the ionization continuum of a sodium cluster located at the center of the fullerene cavity. The strength of this plasmonic RICD, emerging at the otherwise mundane XUV spectrum of isolated $\na$, is of the same order of the cluster's native plasmon resonance, excitable only by the visible light. The study addresses calculations using a cluster-dimer in a spherical orientation. But the essence of the result should be extendable for non-spherical, even longitudinal, dimers. It may further be generalized for polymers, thin films and in the liquid phase. A possible ramification is that higher wavefunction overlaps from a favorable geometry will enhance the effect. Even though experiments will be challenging at the moment, the effect predicted is fundamental, opens a new direction to push the frontier of ICD research, and promises controls in transferring large amount (in plasmonic quantities) of energy to induce remote spectroscopic events.  

\begin{acknowledgments} 
We thank Dr.\ Alexey Popov for encouraging discussions on synthesis possibilities of cluster-fullerenes for future experiments. The research is supported by the SERB, India (HRV), and by the US National Science Foundation Grant Nos.\ PHY-1806206 (HSC), PHY-2110318 (HSC), and CNS-1624416 (the Bartik HPC system, Northwest Missouri State University).
\end{acknowledgments}


\end{document}